\documentclass[apl,reprint,showpacs,aip]{revtex4-1}
\usepackage {graphicx,amsmath}
\usepackage{epstopdf}
\usepackage{bm}
\usepackage{dcolumn}
\begin{document}

\title{Origin of giant magnetoresistance across the martensitic transformation for Ni$ _{44} $Cu$ _{2} $Mn$ _{43} $In$ _{11} $ alloy: Formation of phase fraction}
\author{Mayukh K. Ray$ ^{1} $\footnote{email:mayukh.ray@saha.ac.in}, B. Maji$ ^{2} $, M. Modak$^{1}$ K. Bagani$ ^{1} $, S. Banerjee$ ^{1} $\footnote{email:sangam.banerjee@saha.ac.in}}

\address{$ ^{1} $Surface Physics and Material Science Division, Saha Institute of Nuclear Physics, 1/AF Bidhannagar, Kolkata-700064, India\\$ ^{2} $Acheriya Jagadish Chandra Bose College, 1/1B, AJC Bose Road, Kolkata 700020, India}

\begin{abstract}
We have studied the phase volume fraction related magnetoresistance (MR) across the first order martensite transformation (MT) of Ni$ _{44} $Cu$ _{2} $Mn$ _{43} $In$ _{11} $ alloy. Within the metastability of MT, an isothermal application of magnetic field converts the martensite into austenite. The field induced austenite phase fraction ($f_{IA}$) at any temperature depends on the availability and instability of martensite phase fraction ($f_{M}$) at that temperature. This $f_{IA}$ is found to contribute most significantly to the observed giant MR while the contribution from pure martensite and austenite phase fraction is negligible. It is found that the net MR  follows a non linear proportional relation with the $f_{IA}$ and the ascending and descending branch of $f_{IA}$ follows different power law giving rise to hysteresis in MR. Here we present a detail explanation of the observed behaviours of MR based on the existing phase fraction.
\end{abstract}

\maketitle

The change in electronic and magnetic structure near the martensite transformation (MT) in Ni-Mn based nonstoichiometric Heusler alloys lead to many interesting magnetic and transport properties.\cite{Khan,Oikawa,Krenke,Sharma,JAP,EPL} The MT in these alloys can be tuned by external parameters such as temperature, pressure and magnetic field.\cite{Oikawa,Krenke,SharmaVK} The presence of disorder in the parent phase makes the  MT  occur over a range of these individual control parameter instead of at a sharp value; this range is called the transition width (TW). Thus, within the TW of MT both the austenite and martensite phase coexist in a metastable state. 
A giant magnetoresistance (GMR) is generally observed across the MT just like in multilayer systems, though the later has a different origin for its large MR.\cite{Sharma,JM,JPD} Therefore, it is important to understand in detail  the large magneto-resistive behaviour of these alloys across the MT as a function of applied field. To do so one needs to have knowledge on the existing  volume faction of different phases at each temperature within the MT region when a field is applied or vice-versa. In this context it is noteworthy that the resistivity measurement is more convenient and reliable method over the magnetization measurement \cite{Zuo,Wang,VV1} because for weakly coupled magneto-structural MT, magnetization does not show pronounce anomaly across the MT.\cite{Segui} 
With this motivation, we have focused on the resistivity behaviour of  Ni$ _{44} $Cu$ _{2} $Mn$ _{43} $In$ _{11} $ alloy which shows  MT from ferromagnetic austenite to nearly paramagnetic martensite around a temperature ($T$=) 270 K. We have carried out the calculation of phase volume fraction for martensite ($f_{M}(T)$), total austenite ($ f_{A}(T) $) and induced austenite  ($f_{IA}(T)$) across the TW of MT  under different applied field. It is worth to be noted that the $f_{A}(T)$ is the sum of $f_{IA}(T)$ and pure austenite volume fraction ($f_{PA}(T)$) and both the $f_{IA}$ and $f_{PA}$ has same physical properties as $f_{A}$. Now, it is fascinating to observe that the MR for a given field change ($\Delta H$) attained maximum only when the $f_{IA}$ becomes maximum for the respective $\Delta H$ and the MR does not depend much either on $f_{M}$ or $f_{PA}$. We also found that the MR does not hold a linear relation with the $f_{IA}$ and it's variation  is not the same for the ascending and descending branch of  $f_{IA}$. It should be note it down that, our observation is applicable to all martensite Heusler alloys. The preparation of polycrystalline Ni$ _{44} $Cu$ _{2} $Mn$ _{43} $In$ _{11} $ alloy and the measurements of resistivity is same as our earlier report.\cite{JPD} 

\begin{figure} 
\centering
\includegraphics[scale=0.25]{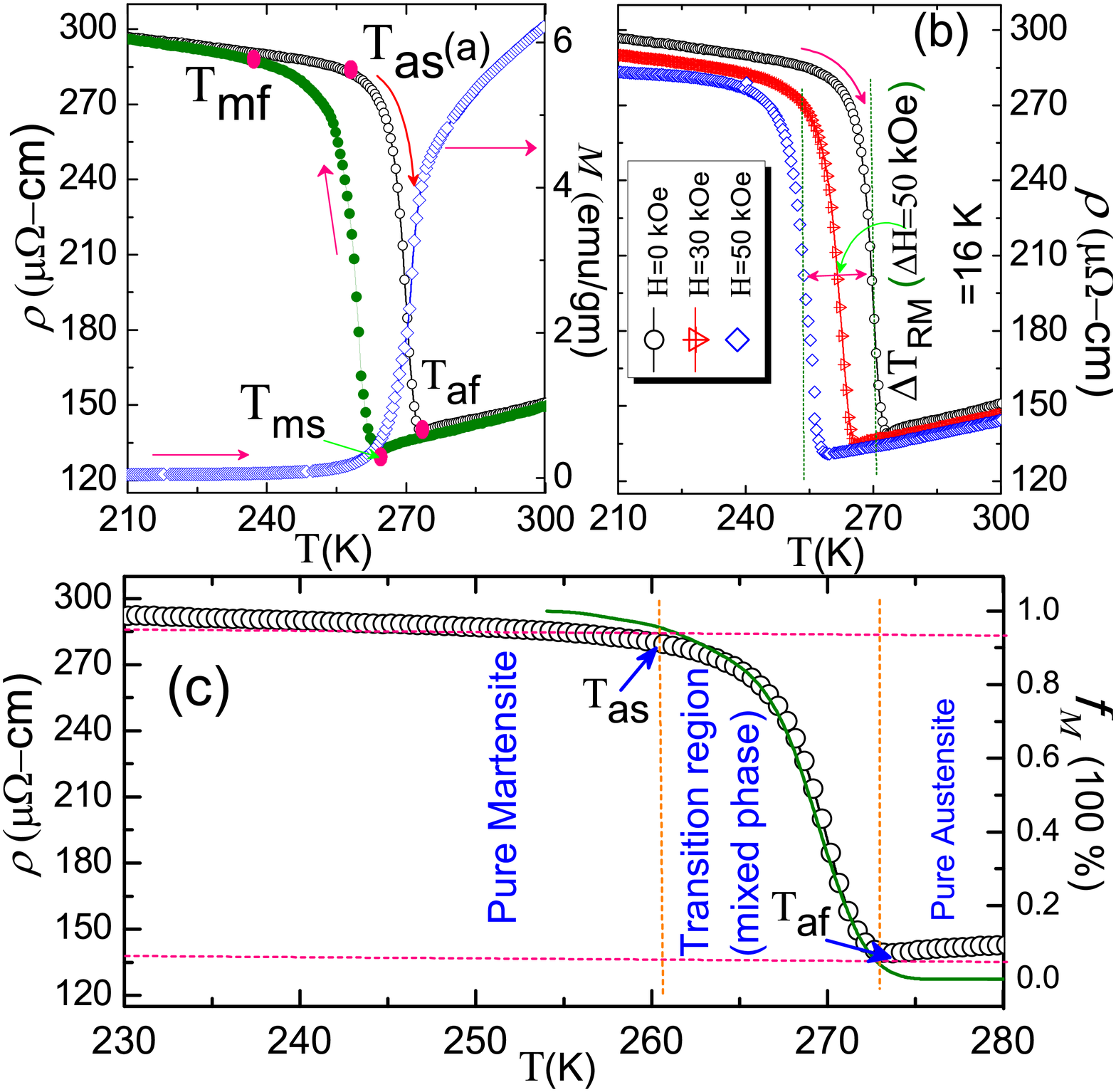}
\caption{(a) The  temperature dependency of $\rho(T)$ and $M(T)$ curve using ZFCC and ZFCW protocol (b) FCW $\rho(T)$ curves at different applied field (c) pink dashed line is the linear extrapolation of $\rho$ curve at $T_{as}$ and $T_{af}$.}
\label{FCC}
\end{figure}
The temperature dependent resistivity ($\rho$) data employing the zero field cooled cooling (ZFCC) and zero field cooled warming (ZFCW) protocols and the thermo-magnetization ($M(T)$) curves obtained using the ZFCW protocol at a field ($H$=)100 Oe are shown in the fig.\ref{FCC}(a). In the ZFCC curve,  resistivity of ferromagnetic austenite phase start  to decrease with the decrease of temperature until the martensite start temperature ($T_{ms}$). This observation reveals, the resistivity of austenite phase follows the metallic behaviour (i.e. $\frac{\partial\rho}{\partial T}>0$) in this temperature region. Below $T_{ms}$, resistivity increases rapidly upto the martensite finish temperature ($T_{mf}$) due to the appearance of ferromagnetic-to-paramagnetic type martensite transformation (MT). This increase in resistivity is partly associated with the enhanced conduction electrons scattering due to development of differently oriented martensite variants along with the other reasonable factors e.g. presence of microcrack, modification of Brillouin zone boundary during MT.\cite{JM,Nayek,VK,VV} But, below $T_{mf}$ the resistivity remains nearly temperature independent. Now during the warming cycle, the resistivity starts to drop at the austenite start temperature ($T_{as}$) and continues to drop upto the austenite finish temperature ($T_{af}$)  after which it follows the same temperature dependence as its ZFCC cycle. The thermal hysteresis observed around the MT region is a manifestation of disorder induced (metastability) first order nature of MT. Being a disorder influenced first order phase transition, the region of thermal hysteresis is highly metastable and both the austenite and martensite phase can coexist there. Since, our motivation is to relate the existing phase fraction with the magnetoresistance ($ \Delta\rho/\rho_{0} $) across the MT, we have measured the $\rho(T)$ at different applied field using the field cooled warming (FCW) protocol where system is cooled under constant field and $\rho$ is measured during the warming cycle at the same field (fig.1(b)). The reverse martensite temperature ($T_{RM}$) is calculated by drawing vertical line which divide the TW into two equal parts. It is evident that with the application of field $ T_{RM} $ is shifted towards  lower temperatures. This effect is the manifestation of the fact that when a magnetic field is applied in the martensite phase at a temperature close to $T_{as}$, the martensite structure transform to austenite at that temperature. Thus, an application of field around or across the MT has a significant role in altering the phase volume fractions in isothermal condition.  A shift ($\Delta T_{RM}$) of about 16 K is observed for application of field of 50 kOe and this high $\Delta T_{RM}$  is a prerequisite condition to achieve a large MR across the MT and  we would also like to address this issue in this article.
\par
The net  $\rho$  at any temperature within the phase coexisting region can always be written as the sum of $ f_{M}(T)\rho_{M} $ and $f_{A}(T)\rho_{A}$ i.e. $\rho(T)$=$f_{M}(T)\rho_{M}$+$f_{A}(T)\rho_{A}$, where $\rho_{M}$ and $\rho_{A}$ are the resistivity of martensite and austenite phase. Since we also know that $f_{M}+f_{A}=1$ thus equation of $\rho(T)$ can be written in terms of $f_{M}(T)$ as:  
\begin{equation}
f_{M}(T)=\dfrac{\rho(T)-\rho_{A}(T)}{\rho_{M}(T)-\rho_{A}(T)}
\label{EQ1}
\end{equation}
The $\rho_{A}(T)$ and $\rho_{M}(T)$ is determined  from just above $T_{af}$  and below (few Kelvin) $T_{as}$, respectively, where the pure austenite and martensite state exists. It should be noted that we are considering the warming curve of $\rho(T,H)$ to avoid the influence of thermal hysteresis. The normalized $f_{M}(T)$ curve plotted as green solid line is shown in fig.\ref{FCC}(c). The shape of the $f_{M}(T)$ curve follows the experimental $\rho(T)$ curve in the phase transition region very well but it attained a maximum just below  $T_{as}$ as the $\rho_{M}$ value used here is determined at few Kelvin below of $T_{as}$.
\begin{figure} 
\centering
\includegraphics[scale=0.3]{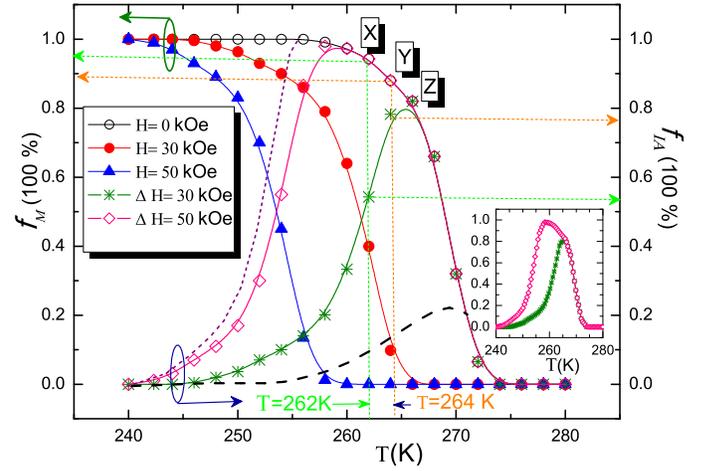}
\caption{The temperature variation of $f_{M}$ and $f_{IA}$ for different applied field and black (thick) and violet (thin) dash line represents schematic $f_{IA}$ curve for $\Delta H$= 10 and 70 kOe, respectively and inset shows the interpolated $f_{IA}$ curves for $\Delta H$=30 and 50 kOe.} 
\label{F2}
\end{figure}
\par
To relate the observed magneto-resistive behaviour with the existing different phases fraction in the transition region one need to quantify the $f_{IA}$ at any temperature under a given applied field apart from the  martensite and austenite phase fraction. It is easy for one to quantify the $f_{M}$ and  $f_{A}$ (=$f_{IA}$+$ f_{PA} $) at any certain temperature on application of different field by measuring the $\rho(T,H)$ curves and transforming them to $f_{M}(T,H)$ curves as shown in fig.\ref{F2}. Since, we are curious to gain a thorough insight into the $f_{IA}$ behaviour on isothermal application of magnetic field, we have used the following formula:
\begin{equation}
f_{IA}(T,\Delta H)=f_{M}(T,0)-f_{M}(T,H)
\label{EQ2}
\end{equation}
where $f_{IA}$ represent the isothermally induced austenite phase fraction for a given field change $\Delta H$. Fig.\ref{F2} shows experimental curves of $f_{IA}$ for $\Delta H$= 30 and 50 kOe. It can be seen that both the curves attain maximum at their respective $T_{af}$ on the $f_{M}(T,H)$ curve and thereafter $f_{IA}$ follows the same path as $f_{M}(T,0)$ curve. Before we get into the detail behaviour of $f_{IA}(T,\Delta H)$ it is convenient to split the whole temperature region of measurement into three parts: (i) $T<T_{as}$ (ii) $T_{as} <T<T_{af}$ (iii) $T>T_{af}$. It is well known from the plot of Landau free energy expression for the first order transition that the martensite and austenite phase is stable in the region (i) and (iii), respectively and in the region (ii) both the phases  become metastable and coexists together.\cite{Chatterjee} When the system is deep inside  the region (i) (say $T$=240 K) the stability of  martensite phase is high, thus the application of field cannot convert  martensite into austenite and $f_{IA}$ become zero. On the other hand as the temperature increases martensite looses its stability and the application of the same field can now convert some fraction of martensite into austenite and hence  $f_{IA}$ starts to increase. When the system just enters into the region (ii), a temperature point say $X$ ($T$=262 K) on the $f_{M}(T,0)$ curve  has $95\%$ martensite phase  ($f_{M}$=95$\%$) and remaining $5\%$ is metastable austenite phase (follow the horizontal green dash line towards the $f_{M}$ axis). Since, the instability of martensite in this region is more than the instability it has in the region (i) thus an application of same field converts $52\%$ of martensite  into austenite ($f_{IA}$=52 $\%$, follow the horizontal green dash line towards the $f_{IA}$ axis) with remaining $f_{M}$=43$\%$ unchanged  and $f_{IA}$ starts to increase rapidly with the progress of temperature. Though the point $Y$ ($T$=264 K) inside the region (ii) contained less volume fraction of martensite [($f_{M})_{Y}<(f_{M})_{X}$] but due to the increased instability of martensite phase, the same field produces $77\%$ austenite ($f_{IA}$=77$\%$) from available $88\%$ martensite with remaining $f_{M}=11\%$ unchanged. Interestingly all the metastable martensite is converted to the austenite  and  $f_{IA}$ reaches maximum (82$\%$) at the point $Z$ ($T$=265 K). Though the martensite instability increases with temperature but due to rapid decrease in available $f_{M}$ leads to rapid drop in $f_{IA}$. The $f_{IA}$ becomes zero above the $T_{af}$ of  $f_{M}(T,0)$ curve as the region beyond the $T_{af}$ belongs to the pure austenite phase thus no martensite phase is available ($f_{M}$=0) which is to be converted into austenite. The same is true for the $f_{IA}$ curve at $\Delta H$=50 kOe. It is  interesting  that the both $f_{IA}$ curves exactly follow $f_{M}(T,0)$ curve during their decrease and this observation is easy to understand from  eq.\ref{EQ2}. One can find that the contribution of $f_{M}(T,H)$ term in eq.\ref{EQ2} becomes zero just beyond the temperature  where $f_{IA}$ reaches its maximum (see the $f_{M}$ curve for respective field) i.e. when $f_{M}(T,H)=0$, $f_{IA}$=$ f_{M}(T,0) $. 
Since, we have observed a nearly 100$\%$ induction of austenite phase by the application of $H$=50 kOe it is thus expected that application of further field will find little influence on the $f_{IA}$.  A schematic expected curve for $\Delta H$= 10 and 70 kOe is shown by the black (thick) and violet (thin) dash line, respectively. 
\begin{figure} 
\centering
\includegraphics[scale=0.25]{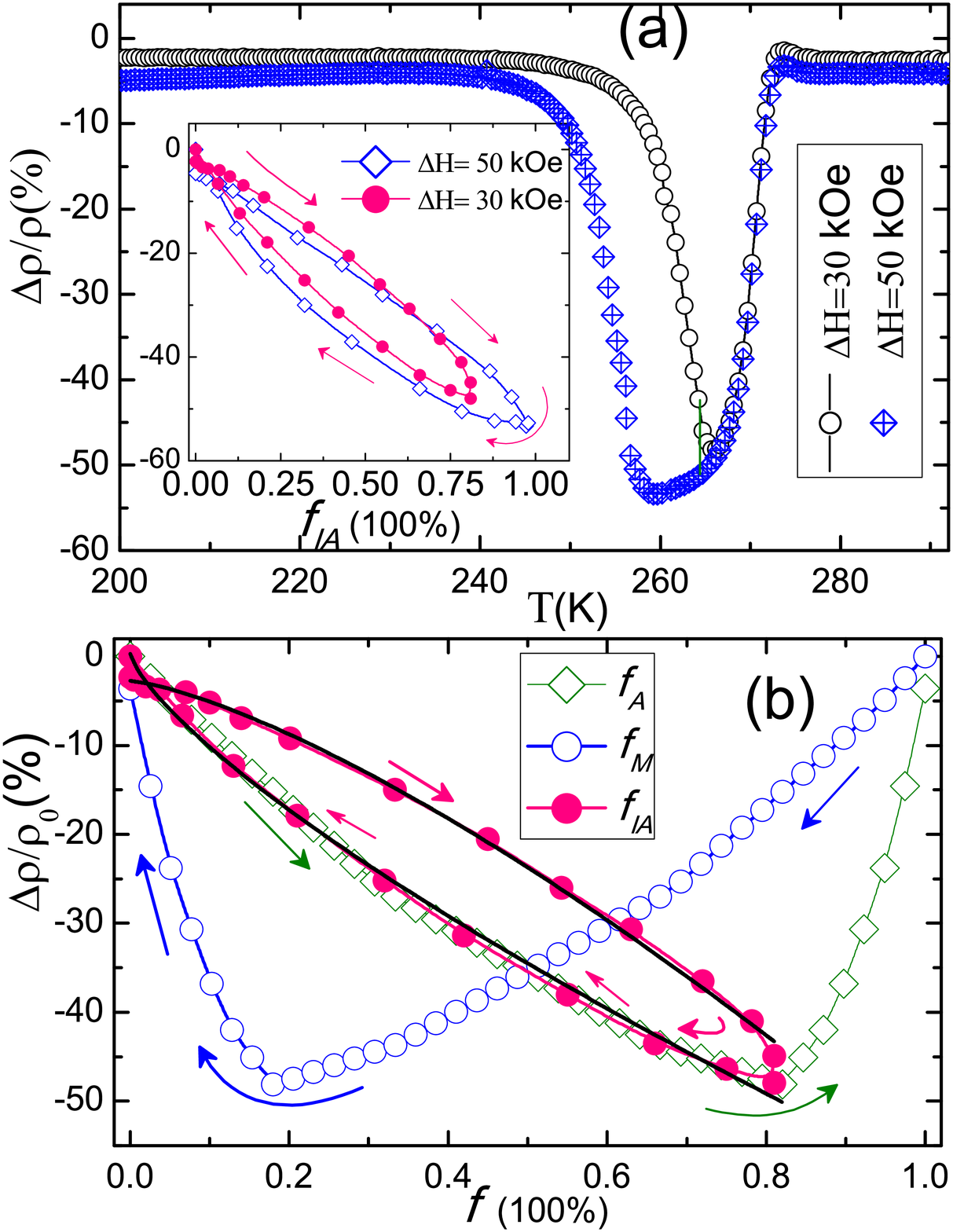}
\caption{(a) Temperature variation of MR for $\Delta H$= 30, 50 kOe and inset shows the variation of MR for a given $\Delta H$ with the $f_{IA}$ calculated at that $\Delta H$ (b) The variation of MR with the $f_{IA}$, $f_{M}$ and $f_{A}$ for $\Delta H$= 30 kOe and the solid curve is obtain after fitting the MR using eq.\ref{EQ3}.}
\label{F3}
\end{figure}
\par
The main panel of fig.\ref{F3} (a) shows the temperature variation of magnetoresistance (MR, $\Delta\rho/\rho_{0}$) for different applied field. The MR is calculated using the formula $\Delta\rho/\rho_{0}=(\rho(T,H)-\rho(T,0))/\rho(T,0$)). It is evident that a large MR is only observed around the phase transformation region and becomes maximum at temperatures 265K and 259K for $\Delta H$=30 and 50 kOe, respectively. Now, invoking our previous arguments i.e the large $\Delta T_{RM}$ under an applied field is an essential criteria to obtain a large MR across the transition region can be understood by expressing the MR in terms of change in resistivity with temperatures ($\frac{d\rho}{dT}$) across the phase transition and the field induced shift of transformation temperature ($\frac{dT}{dH}$) due to the metastability related to the first order MT. Taking into account the above mentioned fact one can approximate the  MR as MR $\propto(\frac{d\rho}{dH})=(\frac{d\rho}{dT})(\frac{dT}{dH})$. Thus, to observe a huge MR, both the factors have to be large. Since, the phase transformation in this type of alloy is always associated with large $\frac{d\rho}{dT}$ due to development of martensite variants which acts as a  scattering centre for conduction electrons and a large $\Delta T_{RM}$ is also observed under the applied field. Thus, the large value of both the factors are mutually producing a giant MR as large as 49$\%$ and 54$\%$ for the $\Delta H$= 30 and 50 kOe, respectively. 
The variation of MR with the $f_{IA}$ for different $\Delta H$ (30 and 50 kOe) is shown in the inset of fig.\ref{F3}(a). It can be seen that, with the increase of $f_{IA}$ and temperature (follow the arrow's direction), MR starts to increase and become maximum when the  $f_{IA}$ attain its maximum. Further increase of temperature leads to the reduction of $f_{IA}$ and MR decreases. It is evident that the MR for the ascending and descending branch of $f_{IA}$  follows  different paths  producing a hysteresis in MR. This irreversibility in MR for the ascending and descending branches of $f_{IA}$ can be understood by recalling the fig.\ref{F2}, i.e. in that figure one can find that for the ascending branch of $f_{IA}$ the total phase fraction consists of the large fraction of martensite, induced austenite and little fraction of pure austenite while on the descending branch of $f_{IA}$ total phase fraction is the sum of large fraction of pure austenite, induced austenite and little amount of martensite, thus it is expected that resistivity and hence the MR would be different for two different branch of $f_{IA}$ due to presence of different amount of phases fraction. But, the fact which is more interesting is that MR does not hold a linear relationship with the $f_{IA}$ instead it seems to follow a certain power law. To know the MR dependence on $f_{IA}$ we have fitted the ascending and descending branch of $f_{IA}$ with MR using the formula
 \begin{equation}
 \frac{\Delta\rho}{\rho_{0}}=- \alpha f_{IA}^{n}
 \label{EQ3}
 \end{equation}
where $\alpha$ defines the strength of MR. The obtained value of $\alpha$ and $n$ for the ascending (superlinear) and descending (sublinear) branch of $f_{IA}$  are listed in the table \ref{Table 1}. It is found that  the value of $n$ for the ascending and descending branch of $f_{IA}$ at $\Delta H$=30 kOe is little higher than its value obtained for $ \Delta H $=50 kOe.  
\begin{table}
\caption{\label{Table 1}
 The parameters obtained from the fitting of MR vs $f_{IA}$ curve using eq.\ref{EQ3}.} 
  \begin{ruledtabular}
  \begin{tabular}{ccc}
  $ \Delta H $(kOe) & $\alpha$  & $n$ \\ 
  \hline
  30 (ascending)& 54$\pm$0.9 & 1.36$\pm$0.4\\ 
  \hline
  30 (descending)& 58$\pm$1.2 & 0.74$\pm$0.3 \\ 
  \hline
  50 (ascending)& 48$\pm$0.9 & 1.24$\pm$0.06 \\ 
  \hline
  50 (descending)& 58$\pm$0.7 & 0.6$\pm$0.04 \\ 
  \end{tabular}
  \end{ruledtabular}
  \end{table}
The fig.\ref{F3}(b) shows the variation of net MR with the $f_{IA}$, $f_{M} $ and $f_{A}$ for $\Delta H$=30 kOe. It can be seen that as the temperature is increasing (increasing direction of temperature is shown by the arrow), $f_{M}$ starts to decrease while $f_{A}$ increases due the increase of $f_{IA}$  and the MR start to increase and  attain maximum when the $f_{IA}$ become maximum. Thereafter with the further increase of temperature, MR  start to decrease though the $f_{A}$ increase ($f_{M}$ decreases). It is noteworthy that the decrease of MR started at where the $f_{IA}$ start to decrease though the $f_{A}$ continue to increases due to the increased contribution from another temperature dependent factor $f_{PA}$.  Thus, combining this with our previous observation, we  argue that the $f_{IA}$ is the most important factor which needs to be considered while discussing the giant MR across the MT for this type of alloys because the resistivity of pure martensite or austenite phase does not change significantly. This mechanism can also be observed GMR in magnetic oxide such as doped manganates where field induced magnetic transition is observed with structural transformation.\cite{Rao} 
In conclusion, we have presented a detail estimation of various phase fraction at  any given temperature and field, in particular the martensitic phase transition region, together with a detail analysis of the corresponding  MR data. Our results, reveals that the  $f_{IA}$ has major contribution for giant MR than the other factors though the $f_{IA}$ at any particular temperature  depends on the availability and instability of $f_{M}$ at that temperature. The MR  varies proportionally with the $f_{IA}$ but the relation is not linear. The ascending and descending  branch of $f_{IA}$ does not follow the same power law, thus giving rise to hysteresis in MR.
One of the authors MKR like to thank material science division of VECC, Kolkata for XRD facility and UGC for financial assistance.


\begin{thebibliography}{99}
\bibitem{Khan} M. Khan, I. Dubenko, S. Stadler and Naushad Ali, Appl. Phys. Lett.\textbf{91}, 072510 (2007).

\bibitem{Oikawa} K. Oikawa,W.Ito,Y. Imano, R. Kanuima, K. Ishida, S. Okamoto, O.Kitakami and T. Kanomata, Appl. Phys. Lett.\textbf{88}, 122507 (2006).

\bibitem{Krenke} T. keneke, M.Acet, E.F Wassermann, X. Moya, L. Manosa and A. Planes, Phys. Rev. B \textbf{73}, 173413 (2006).

\bibitem{Sharma} V. K. Sharma, M. K. Chattophadhyay,K. H. B. Saheb, Anil Chouhan and S. B. Roy, Appl. Phys. Lett. \textbf{89}, 222509 (2007).

\bibitem{JAP} M. K. Ray, K. Bagani, R. K. Singh, B. Majumdar and S. Banerjee, J. Appl. Phys. \textbf{114}, 123904 (2013).

\bibitem{EPL} M. K. Ray, K. Bagani, P. K. Mukhopadhyay and S. Banerjee, Euro. Phys. Lett. \textbf{109}, 47006 (2015).

\bibitem{SharmaVK} V. K. Sharma, M. K. Chattophadhyay and S. B. Roy, Phys. Rev. B  \textbf{76}, 140401(R) (2007).

\bibitem{JM} J. M. Barandiaran, V. A. Chernenko, P. Lazpita, J. Gutierrez and J. Feuchtwanger, Phys. Rev. B \textbf{80}, 104404 (2009).

\bibitem{JPD} M. K. Ray, B. Maji, K. Bagani and S. Banerjee, J. Phys.D: Appl. Phys. \textbf{47}, 385001 (2014).

\bibitem{Zuo} F. Zuo, X. Su and K. H Wu, Phys. Rev. B \textbf{58}, 11127 (1998).

\bibitem{Wang} W. H. Wang, J. L. Chen, S. X. Gao, G. H. Wu, Z. Wang, Y. F. Zhen, L. C. Zhao and W. S. Zhan, J. Phys.: Condens Matter. \textbf{13}, 2607 (2001).

\bibitem{VV1} V. V. Khovalio, V. Novosad, T. Takagi, D. A. Filippov, R. Z. Levitin and A. N. Vasil's, Phys. Rev. B \textbf{70}, 174413 (2004).

\bibitem{Segui} C. Segui, V. A. Chernenko, J. Pons, E. Cesari, V. V. Khovalio and T. Takagi, Acta. Mater. \textbf{53}, 111 (2005).

\bibitem{Nayek} A. K. Nayak,K. G. Suresh and A. K. Nigam, Acta. Mater. \textbf{59}, 3304 (2011).

\bibitem{VK} V. K. Sharma, M. K. Chattopadhyay and S.B. Roy, Phys. Rev. B \textbf{76}, 140401(R) (2007).

\bibitem{VV} V. V. Khovaylo, T. Omori, E. Endo, X. Xu, R. Kainuma, A. P. Kazakov, V. N. Prudnikov, E. A. Ganshina, A.I. Novikov, Yu. O. Mikhailovsky, D.E. Mettus and A. B. Granovsky, Phys. Rev. B \textbf{87}, 174410 (2013).

\bibitem{Chatterjee} S. Chatterjee, S. Giri and S. Majumdar, Phys. Rev. B \textbf{77}, 012404 (2008).

\bibitem{Rao} R. Seshadri, C. Martin, M. Hervieu, and B. Raveau and C. N. R. Rao, Chem. Mater. \textbf{9}, 270 (1997).












 
\end{thebibliography}
\end{document}